\documentclass[twoside,fleqn]{article}
\usepackage{npb,epsfig}
\newcommand{\AmS}{{\protect\the\textfont2
  A\kern-.1667em\lower.5ex\hbox{M}\kern-.125emS}}

\def\beq{\begin{equation}}
\def\eeq{\end{equation}}
\def\beqar{\begin{eqnarray}}
\def\eeqar{\end{eqnarray}}
\def\barr#1{\begin{array}{#1}}
\def\earr{\end{array}}
\def\bfi{\begin{figure}}
\def\efi{\end{figure}}
\def\btab{\begin{table}}
\def\etab{\end{table}}
\def\bce{\begin{center}}
\def\ece{\end{center}}

\def\text{\textstyle}

\def\al{\alpha}
\def\be{\beta}

\def\de{\delta}

\def\si{\sigma}

\def\De{\Delta}

\def\reffi#1{\mbox{Fig.~\ref{#1}}}

\def\citere#1{\mbox{Ref.~\cite{#1}}}

\def\mathswitchr#1{\relax\ifmmode{\mathrm{#1}}\else$\mathrm{#1}$\fi}

\newcommand{\PW}{\mathswitchr W}
\newcommand{\PZ}{\mathswitchr Z}
\newcommand{\PA}{\mathswitchr A}

\newcommand{\PH}{\mathswitchr H}

\newcommand{\Ph}{\mathswitchr h}

\newcommand{\Pb}{\mathswitchr b}

\newcommand{\Pt}{\mathswitchr t}

\def\mathswitch#1{\relax\ifmmode#1\else$#1$\fi}

\newcommand{\MW}{\mathswitch {M_\PW}}

\newcommand{\MZ}{\mathswitch {M_\PZ}}
\newcommand{\MH}{\mathswitch {M_\PH}}

\newcommand{\Mt}{\mathswitch {m_\Pt}}

\newcommand{\mh}{\mathswitch {m_\Ph}}

\newcommand{\MA}{\mathswitch {M_\PA}}

\newcommand{\sweff}{\sin^2 \theta_{\mathrm{eff}}}

\def\tb{\tan\beta}

\newcommand{\mste}{m_{\tilde{\Pt}_1}}
\newcommand{\mstz}{m_{\tilde{\Pt}_2}}

\newcommand{\mt}{\Mt}

\newcommand{\mgl}{m_{\tilde{\mathrm{g}}}}

\newcommand{\Stop}{\tilde{\Pt}}

\newcommand{\Stopz}{\tilde{\Pt}_2}

\newcommand{\tst}{\theta_{\tilde{\Pt}}}

\newcommand{\tsf}{\theta\kern-.20em_{\tilde{f}}}
\newcommand{\tsfp}{\theta\kern-.20em_{\tilde{f}\prime}}
\newcommand{\tsq}{\theta\kern-.15em_{\tilde{q}}}

\newcommand{\sintt}{\sin\tst}

\newcommand{\msusy}{M_{\mathrm{SUSY}}}

\newcommand{\lsim}
{\;\raisebox{-.3em}{$\stackrel{\displaystyle <}{\sim}$}\;}
\newcommand{\gsim}
{\;\raisebox{-.3em}{$\stackrel{\displaystyle >}{\sim}$}\;}

\newcommand{\feh}{{\em FeynHiggs}}

\newcommand{\cp}{{\cal CP}}

\newcommand{\onel}{one-loop}

\newcommand{\mhmax}{\mh^{\mathrm{max}}}

\newcommand{\VL}{\left( \begin{array}{c}}
\newcommand{\VR}{\end{array} \right)}
\newcommand{\ML}{\left( \begin{array}{cc}}
\newcommand{\MLd}{\left( \begin{array}{ccc}}
\newcommand{\MLv}{\left( \begin{array}{cccc}}
\newcommand{\MR}{\end{array} \right)}

\newcommand{\gev}{\,\, \mathrm{GeV}}

\newcommand{\BC}{\begin{center}}
\newcommand{\EC}{\end{center}}
\newcommand{\BE}{\begin{equation}}
\newcommand{\EE}{\end{equation}}
\newcommand{\BEA}{\begin{eqnarray}}
\newcommand{\BEAnn}{\begin{eqnarray*}}
\newcommand{\EEA}{\end{eqnarray}}
\newcommand{\EEAnn}{\end{eqnarray*}}

\newcommand{\id}{{\rm 1\kern-.12em
\rule{0.3pt}{1.5ex}\raisebox{0.0ex}{\rule{0.1em}{0.3pt}}}}

\def\draftdate{\relax}
\def\mda{\relax}
\def\mua{\relax}
\def\mla{\relax}
\def\draft{
\def\thtystars{******************************}
\def\sixtystars{\thtystars\thtystars}
\typeout{}
\typeout{\sixtystars**}
\typeout{* Draft mode!
         For final version remove \protect\draft\space in source file
*}
\typeout{\sixtystars**}
\typeout{}
\def\draftdate{\today}
\def\mua{\marginpar[\boldmath\hfil$\uparrow$]%
                   {\boldmath$\uparrow$\hfil}%
                    \typeout{marginpar: $\uparrow$}\ignorespaces}
\def\mda{\marginpar[\boldmath\hfil$\downarrow$]%
                   {\boldmath$\downarrow$\hfil}%
                    \typeout{marginpar: $\downarrow$}\ignorespaces}
\def\mla{\marginpar[\boldmath\hfil$\rightarrow$]%
                   {\boldmath$\leftarrow $\hfil}%
                    \typeout{marginpar:
$\leftrightarrow$}\ignorespaces}
\def\Mua{\marginpar[\boldmath\hfil$\Uparrow$]%
                   {\boldmath$\Uparrow$\hfil}%
                    \typeout{marginpar: $\Uparrow$}\ignorespaces}
\def\Mda{\marginpar[\boldmath\hfil$\Downarrow$]%
                   {\boldmath$\Downarrow$\hfil}%
                    \typeout{marginpar: $\Downarrow$}\ignorespaces}
\def\Mla{\marginpar[\boldmath\hfil$\Rightarrow$]%
                   {\boldmath$\Leftarrow $\hfil}%
                    \typeout{marginpar:
$\Leftrightarrow$}\ignorespaces}
\overfullrule 5pt
\oddsidemargin -15mm
\marginparwidth 29mm
}

\title{Higgs-mass predictions and electroweak precision observables in
       the Standard Model and the MSSM}

\author{S.~Heinemeyer\address{DESY Theorie, Notkestr.~85,
        D--22603 Hamburg, Germany} 
        and
        G.~Weiglein\address{CERN, TH Division, CH--1211 Geneva 23,
        Switzerland}}

\begin{document}

\thispagestyle{empty}

\def\thefootnote{\fnsymbol{footnote}}

\begin{flushright}
CERN--TH/2000--225\\
DESY 00--108\\
hep-ph/0007307\\
\end{flushright}

\vspace{1cm}

\begin{center}

{\Large\sc {\bf Higgs-mass predictions and electroweak precision 
observables\\[.5em]
in the Standard Model and the MSSM%
\footnote{To appear in the proceedings of the {\em 5th Zeuthen Workshop
on Elementary Particle Theory ``Loops and Legs in Quantum Field
Theory''}, Bastei/K\"onigstein, Germany, April 9--14, 2000.}
}}\\[3.5em]
{\large
{\sc
S.~Heinemeyer$^{1}$ and G.~Weiglein$^{2}$%
}
}

\vspace*{1cm}

{\sl
$^1$ DESY Theorie, Notkestr.~85, D--22603 Hamburg, Germany

\vspace*{0.4cm}

$^2$ CERN, TH Division, CH-1211 Geneva 23, Switzerland
}

\end{center} 

\vskip 5cm

Higher-order results for electroweak precision observables in the Standard 
Model are analyzed in view of the experimental accuracies
achievable at the present and the next generation of colliders, and the
indirect prediction of the Higgs-boson mass from the precision data is 
discussed. Within the MSSM, 
two-loop results for the lightest $\cp$-even Higgs-boson mass 
are confronted with the exclusion limits from LEP. Possible precision tests
of the MSSM at a future linear collider are furthermore investigated.
\par

\vfill

\noindent
June 2000
\null

\def\thefootnote{\arabic{footnote}}
\setcounter{page}{0}
\setcounter{footnote}{0}

\clearpage

\begin{abstract}
Higher-order results for electroweak precision observables in the Standard 
Model 
are analyzed in view of the experimental accuracies
achievable at the present and the next generation of colliders, and the
indirect prediction of the Higgs-boson mass 
from the precision data is discussed. Within the MSSM, 
two-loop results for the lightest $\cp$-even Higgs-boson mass 
are confronted with the exclusion limits from LEP. Possible precision tests
of the MSSM at a future linear collider are furthermore investigated.
\end{abstract}

\maketitle

\section{INTRODUCTION}

By confronting the electroweak precision data with the theory, i.e.\ the
electroweak Standard Model (SM) or extensions of it, most prominently
the Minimal Supersymmetric Standard Model (MSSM), it is possible to test 
the theory at its quantum level, where all parameters of the model enter. 
Within the SM, a fit to the precision data allows to constrain the mass
of the Higgs boson, being the last missing ingredient of the model. The
present 95\% C.L.\ upper bound on the Higgs-boson mass is given by
$\MH < 188$~GeV~\cite{mori00}. The dependence of the precision
observables on $\MH$ is only 
logarithmic in leading order, while they depend quadratically
on the top-quark mass~\cite{velt}. Thus, in order to derive
indirect constraints on the Higgs-boson mass a very high precision of
the experimental data and the theoretical predictions is needed. 

While within the SM only the mass of the Higgs boson has not been
experimentally determined so far,
the MSSM in its unconstrained form (i.e.\ without 
specific assumptions about the SUSY-breaking mechanism) introduces more 
than 100 unknown free parameters (masses, mixing angles, etc.).
A precise determination of the model parameters will not
only be important in order to investigate whether the MSSM is consistent
with the data, but also to infer possible patterns of the underlying
SUSY-breaking mechanism.
In this context the indirect constraints on the model from precision data will
often be complementary to the information from the direct production of
SUSY particles. Furthermore, a very stringent direct test of the MSSM 
is possible. In contrast to the SM, the mass of the lightest
$\cp$-even Higgs boson, $\mh$, is not a free parameter in the MSSM but
is calculable from other parameters of the model. This results in
the tree-level bound $\mh < \MZ$, which however is strongly affected by
large radiative corrections~\cite{mhiggs1}, shifting it to about
$\mh \lsim 135$~GeV at the two-loop 
level~\cite{mh2rg1,mh2rg2,mh2rg3,mh2fd,mh2ep}. 
If the lightest $\cp$-even
Higgs boson of the MSSM will be detected, its mass will play an
important role as a precision observable. The prospective accuracy at
the LHC is $\De\mh \approx 0.2$~GeV~\cite{lhctdr}. At a future linear collider 
this could be improved to $\De\mh \approx 0.05$~GeV~\cite{teslatdr}, while 
at a future muon collider even an accuracy of 
$\De\mh \approx 0.1$~MeV~\cite{mumh} could be achievable.

\section{PRECISION OBSERVABLES IN THE SM 
AT PRESENT AND FUTURE COLLIDERS}

The present bound on the Higgs-boson mass derived within the SM by
comparing the experimental data with the theoretical predictions
arises mainly from the precision measurements of the mass of the 
W~boson, $\MW = 80.419 \pm 0.038$~GeV~\cite{mori00}, and the effective
leptonic weak mixing angle at the Z-boson resonance, 
$\sweff = 0.23149 \pm 0.00017$~\cite{mori00}. Theoretical
uncertainties in the predictions for the observables arise both from
unknown higher-order corrections and from the experimental errors of
the input parameters (in particular $\mt$, $\De\al_{\mathrm{had}}$) used 
for the theoretical predictions. 

\begin{figure}[htb]
\centerline{
\psfig{figure=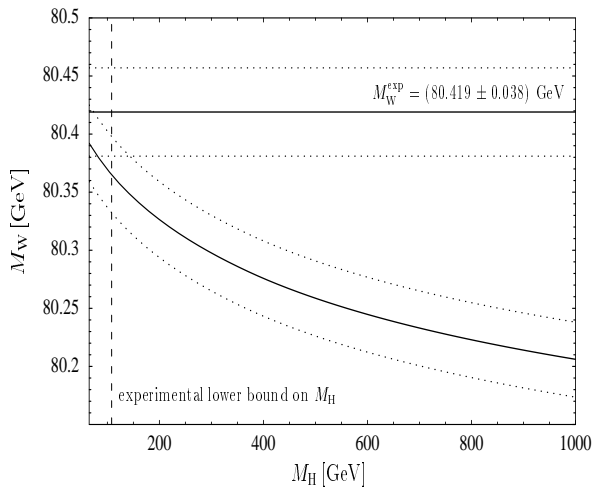,width=8cm,height=6.4cm}
}
\vspace{-6ex}
\caption{The prediction for $\MW$ as a function of $\MH$ for
$\mt = 174.3 \pm 5.1$~GeV is compared with the current experimental
value, $\MW^{\mathrm{exp}} = 80.419 \pm 0.038$~GeV~\cite{mori00}, and
the experimental 95\% C.L.\ lower bound on the Higgs-boson mass,
$\MH = 107.9$~GeV~\cite{lephiggs}.
\label{fig:mw2l}
}
\vspace{-1em}
\end{figure}

The constraints on $\MH$ arising from the prediction for the W-boson
mass are illustrated in \reffi{fig:mw2l}, where the most recent
SM result for $\MW$~\cite{delr} is compared with the current experimental 
value. The present 95\% C.L.\ lower bound on $\MH$ from the direct
search of 
$\MH = 107.9$~GeV~\cite{lephiggs} is also indicated. The plot shows the
well-known preference for a light Higgs boson within the SM. Confronting 
the theoretical prediction (allowing a variation of $\mt$, which at
present dominates the theoretical uncertainty, within $1\sigma$) with
the $1\sigma$ region of $\MW^{\mathrm{exp}}$
and the 95\% C.L.\ lower bound on $\MH$, only 
a rather small region in the plot matches all three constraints.

\begin{figure}[htb]
\centerline{
\psfig{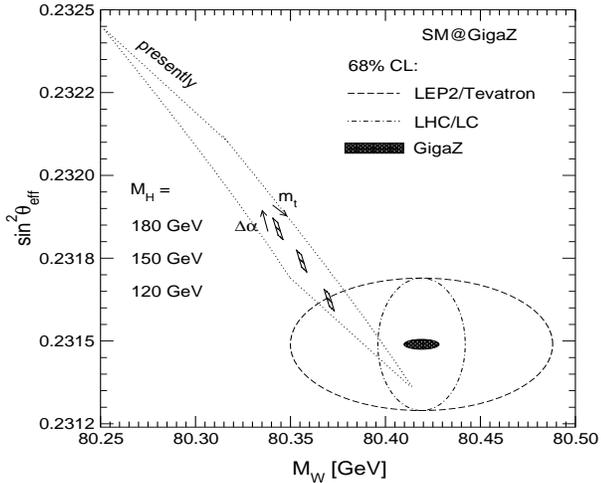}
}
\vspace{-6ex}
\caption{The present and prospective future theoretical predictions in
the SM (for
three hypothetical values of the Higgs-boson mass) are compared with the
experimental accuracies at LEP2/Tevatron (Run IIA), the LHC and GigaZ.
\label{fig:swmwsm}
}
\vspace{-1em}
\end{figure}

At the next generation of colliders a
significant improvement can be expected in the experimental accuracies 
of both the observables employed for testing the theory and of the input
parameters used for deriving the theoretical predictions. This is 
illustrated
in \reffi{fig:swmwsm}, where the SM predictions for $\MW$ and $\sweff$
are compared with the
experimental accuracy (assuming the present experimental central values
of the observables) obtainable at LEP2, SLC and the Tevatron (Run IIA)
as well as with prospective future accuracies at the LHC and at a
high-luminosity linear collider in a dedicated low-energy run
(GigaZ). The experimental accuracies assumed in
\reffi{fig:swmwsm} for LEP2/Tevatron, LHC and GigaZ are $\De \MW =
30$~MeV, 15~MeV, 6~MeV and $\De \sweff = 1.7 \times 10^{-4}$, $1.7
\times 10^{-4}$, $1 \times 10^{-5}$, respectively.
The allowed region of
the present SM prediction corresponds to varying $\MH$ in the 
interval $90 \gev \leq \MH \leq 400 \gev$ and $\mt$ within its current
experimental $1 \sigma$ uncertainty. The theoretical prediction at GigaZ, 
assuming
that the Higgs boson has been found, is shown for three values of the
Higgs-boson mass, $\MH = 120, 150, 180$~GeV, and an uncertainty of 
$\de\mt = \pm 200$~MeV and $\de \De\al = \pm 7 \times 10^{-5}$ is taken
into account in this case. As can be seen in \reffi{fig:swmwsm}, the 
precision observables
$\MW$ and $\sweff$ provide a very sensitive test of the theory, in
particular in the case of the GigaZ accuracy~\cite{sitgesPO,gigaz}.

This fact manifests itself in the precision with which $\MH$ can
indirectly be determined within the SM. While at present the $1\sigma$
uncertainty of $\de \MH/\MH$ derived from all data 
amounts to more than $\pm 60\%$, at GigaZ an accuracy of about 
$\de \MH/\MH = \pm 7\%$ can be expected~\cite{gigaz},%
\footnote{For the future theoretical uncertainties from unknown
higher-order corrections (including the uncertainties from $\de\De\al$)
we have assumed $\delta\MW(\mbox{theory}) = \pm 3$~MeV and
$\delta\sweff(\mbox{theory}) = \pm 3 \times 10^{-5}$~\cite{gigaz}.}
which is of
the same order of magnitude as presently the indirect determination of
$\mt$ from the precision data (without using its experimental value). 

\section{CONSTRAINTS ON $\tan\be$ FROM THE HIGGS SEARCH AT LEP}

The upper bound of about $\mh \lsim 135$~GeV on the mass of the lightest
$\cp$-even Higgs boson is a definite and fairly robust prediction
of the MSSM, which can be tested at the present and the next generation
of colliders. By comparing the experimental limit on $\mh$ from
the search at LEP2 with the theoretical result for the upper bound on
$\mh$ in the MSSM as a function of $\tb$ (the ratio of the vacuum
expectation values of the two Higgs doublets) it is possible to derive 
constraints on $\tb$. 

\begin{figure}[htb]
\centerline{
\psfig{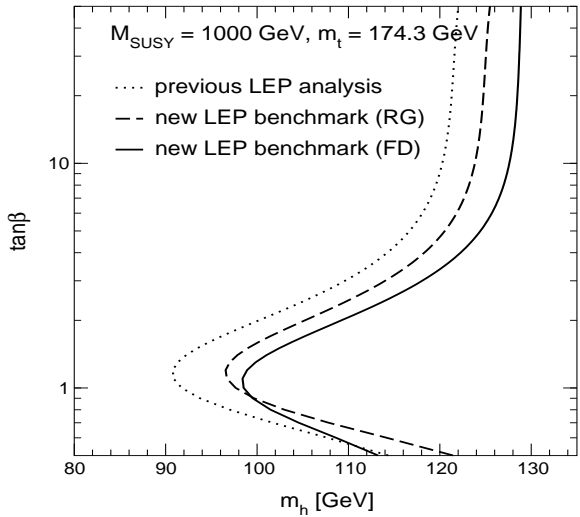}
}
\vspace{-6ex}
\caption{The mass of the lightest $\cp$-even Higgs boson in the MSSM
is shown as a
function of $\tan\be$ using the RG result with the parameter settings
previously used for the LEP analyses, the RG result in the
$\mhmax$~scenario, and the FD result in the $\mhmax$~scenario (see
text).
\label{fig:tbexcl}
}
\vspace{-1em}
\end{figure}

The theoretical input used in the analysis of the search at LEP for the
lightest $\cp$-even Higgs boson of the MSSM has recently been
improved in two ways. Firstly, a new ``benchmark scenario'' has been
proposed~\cite{tbexcl,bench} which slightly generalizes the parameter
settings (for $\mt = 174.3$~GeV and $\msusy = 1$~TeV kept fixed) used so
far in the LEP analyses. It leads to a more conservative upper bound on
$\mh$ as a function of $\tan\be$ (``$\mhmax$~scenario''). Secondly, 
a new Feynman-diagrammatic two-loop result for $\mh$ became
available~\cite{mh2fd} which incorporates genuine two-loop
contributions (see \citere{mhgen2l}) 
that were not contained in the renormalization-group (RG)
improved one-loop effective potential result~\cite{mh2rg2} previously used
for the data analyses. This is illustrated in \reffi{fig:tbexcl}, where the
upper bound on $\mh$ is shown as a function of $\tan\be$.
The difference between the dashed and the dotted curve is the 
effect of changing the previously used parameter settings to the new LEP
benchmark values ($\mhmax$~scenario), while the difference between the
full curve and the dashed curve is caused by  using the
Feynman-diagrammatic (FD) result
instead of the previous result (RG). As can be seen in the figure, both
changes (for $\tan\be \gsim 1$) result in a sizable shift towards 
higher values of $\mh$. Their
combined effect amounts to an upward shift of $\mh$ of up to almost $10$~GeV.

\begin{figure}[ht!]
\vspace{-1em}
\centerline{
\psfig{figure=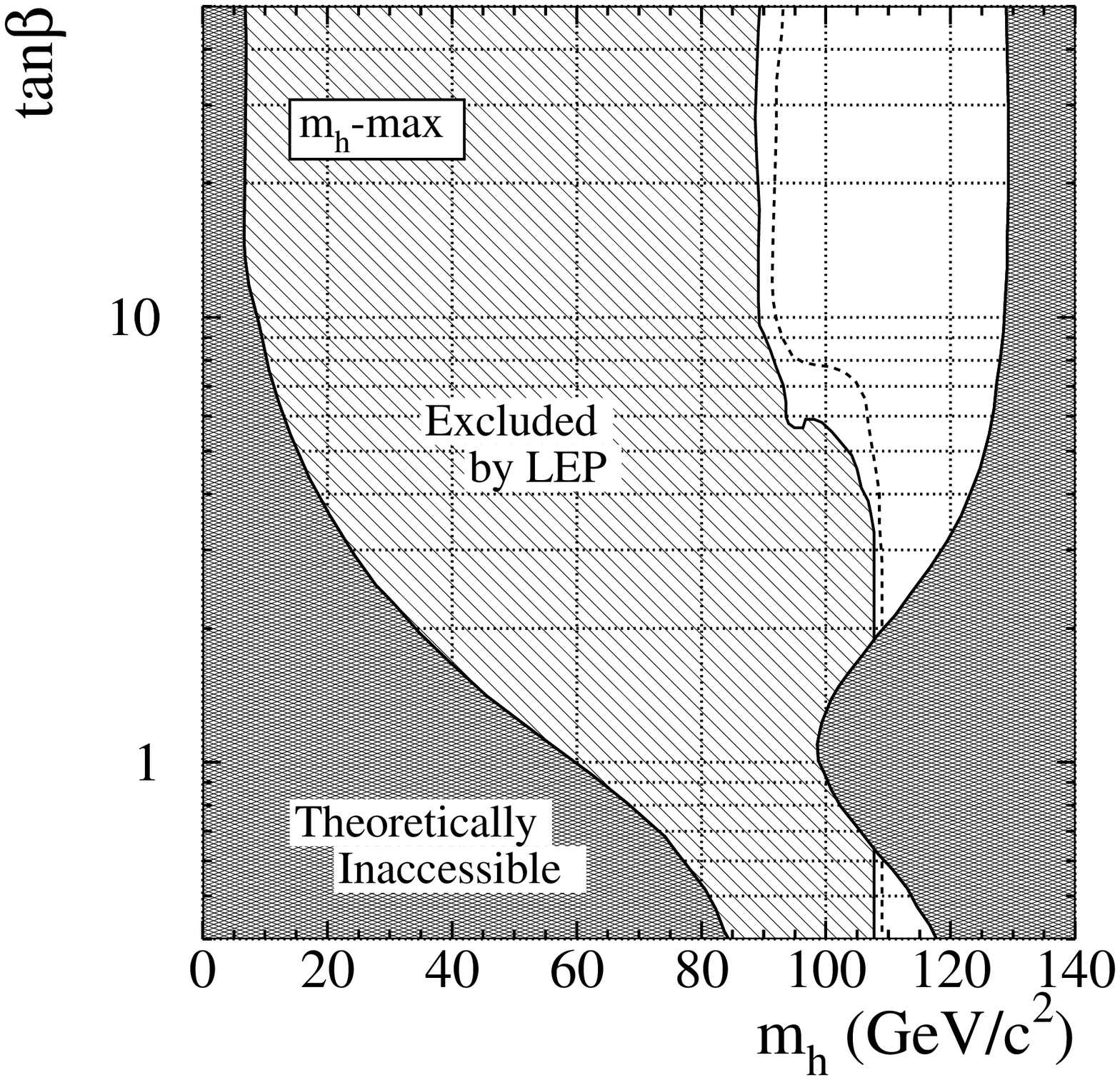,width=8cm,height=6.5cm}
}
\centerline{
\psfig{figure=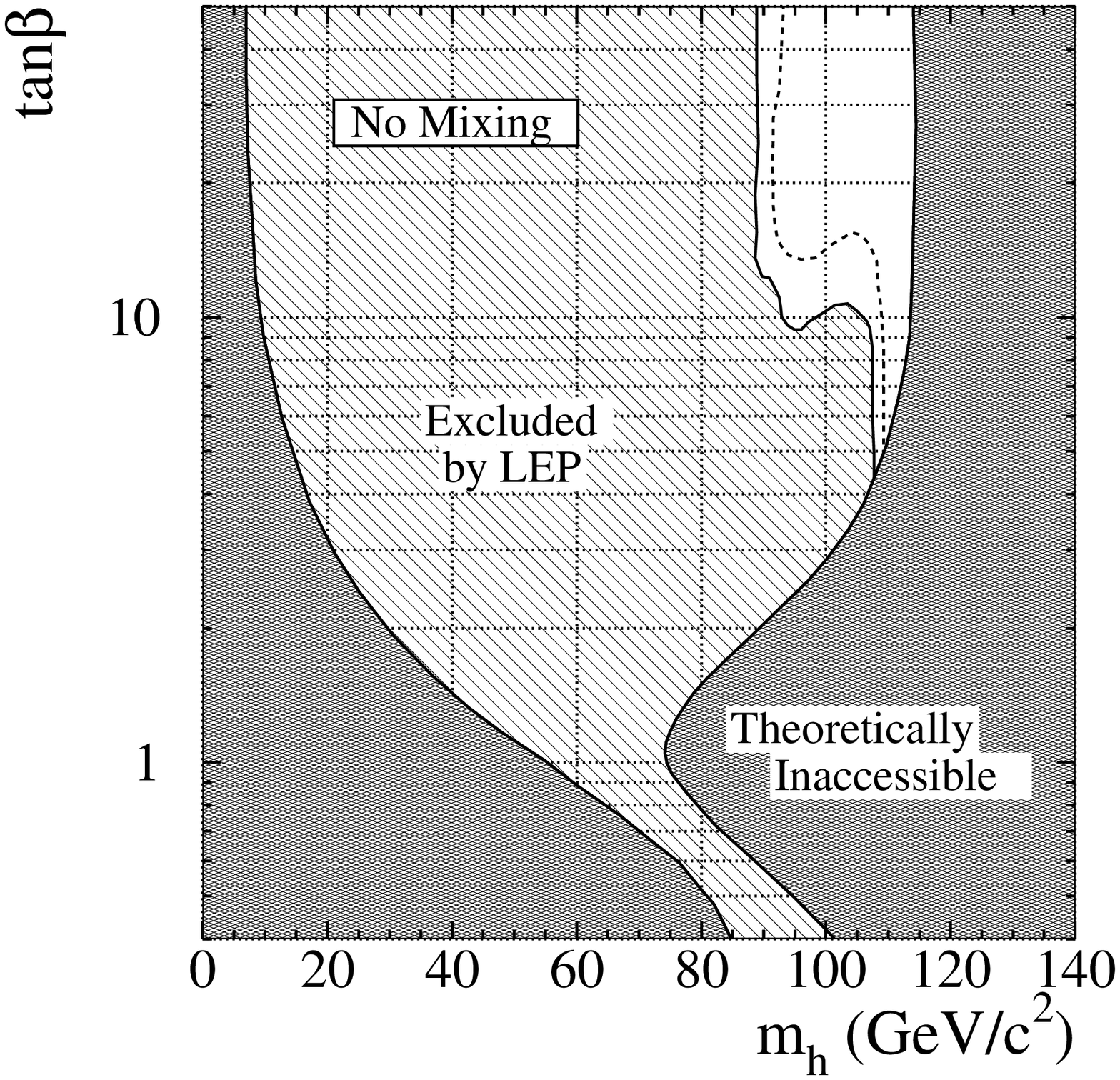,width=8cm,height=6.5cm}
}
\vspace{-6ex}
\caption{The 95\% C.L.\ bounds on $\mh$ in the $\mhmax$ and the no-mixing
scenario obtained from combining the data of
the four LEP experiments at 192 to 202~GeV with earlier data taken at
lower energies are compared with the upper bound on $\mh$ within the 
MSSM~\cite{lephiggs}.
The full lines represent the actual observation, while the dashed lines
are the expected limits based on ``background only'' Monte Carlo
simulations.
\label{fig:mhlep}
}
\vspace{-1em}
\end{figure}

This upward shift in $\mh(\tan\be)$ gives rise to the fact that the
excluded $\tan\be$ region in the $\mhmax$~scenario using the 
combined data taken by the four LEP collaborations up to the 
end of 1999 is smaller than in the previous LEP analyses, despite the fact
that the experimental lower bound on $\mh$ as function of $\tan\be$ 
has considerably improved. In \reffi{fig:mhlep}, the excluded region 
resulting from the combination of the data of the four LEP experiments 
is compared with the upper (and lower) bound within the MSSM (marked as
``theoretically inaccessible'') obtained
with the program \feh~\cite{feynhiggs} based on the Feynman-diagrammatic
result. The upper plot shows the case of the $\mhmax$~scenario, for which
an excluded region of $0.7 < \tan\be < 1.8$ can be
inferred~\cite{lephiggs}. The lower plot shows the so-called no-mixing
scenario~\cite{bench}, where vanishing mixing in the scalar top sector 
is assumed
(the other parameters are the same as in the $\mhmax$~scenario),
resulting in significantly smaller values of $\mh$. In this scenario an
excluded region of $0.4 < \tan\be < 4.1$ is obtained~\cite{lephiggs}.

\section{PRECISION TESTS OF THE MSSM AT GigaZ}

\begin{figure}[htb]
\centerline{
\psfig{figure=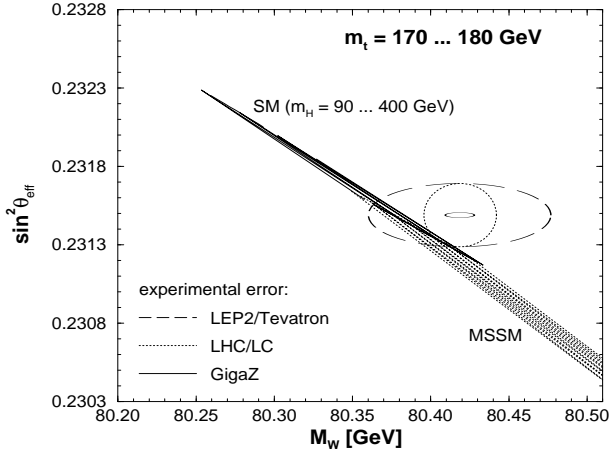,width=8cm,height=6.1cm}
}
\vspace{-6ex}
\caption{Theoretical predictions of the SM and the MSSM in the 
$\MW$--$\sweff$ plane compared with expected experimental accuracies at
LEP2/Tevatron, the LHC and GigaZ.
}
\label{fig:swmwsmmssm}
\vspace{-1em}
\end{figure}

Similarly to the case of the SM, the predictions for $\MW$ and $\sweff$
can also be employed for an indirect test of the MSSM. We consider here
the unconstrained MSSM with real parameters. For the precision
observables and $\mh$ mainly the parameters of the scalar top and bottom
sector and of the Higgs sector are relevant. 
In \reffi{fig:swmwsmmssm} the predictions within the SM and the MSSM
in the $\MW$--$\sweff$ plane are compared with the experimental
accuracies at LEP2/Tevatron, the LHC (and an LC without the GigaZ mode) 
and GigaZ. For the SUSY contributions to $\MW$ and $\sweff$ we use
the  complete \onel\ results in the MSSM~\cite{susyprecOL} combined with
the leading higher-order QCD corrections~\cite{mssm2lqcd}.%
\footnote{The recent
electroweak two-loop results of the SM part in the MSSM~\cite{delr,sm2lnl}
have not been taken into account, since no genuine MSSM counterpart is
available so far.}
The allowed region of the SM prediction is the same as
in \reffi{fig:swmwsm}, while in the MSSM prediction besides the
uncertainty of $\mt$ also the SUSY parameters are varied. 
\reffi{fig:swmwsmmssm}
shows that the predictions of the two models have only a relatively
small overlap in the $\MW$--$\sweff$ plane (corresponding to the SM with 
a light Higgs boson and the MSSM in the decoupling region). With the
GigaZ accuracy, the precision observables provide a high sensitivity to
deviations from both models.

The constraints from the 
precision data on $\MW$,
$\sweff$, etc.\ at GigaZ can be combined with those from a prospective 
measurement of $\mh$ with a precision of $\de \mh = 0.05$~GeV. This
could allow to
indirectly probe the masses of particles in supersymmetric theories 
that might not be accessible directly neither at the LHC nor at the LC.

As an example, we consider the case where at the LHC and the LC
the mass of the lighter scalar top quark, $\mste$, and the
$\Stop$-mixing angle, $\tst$, are measurable very well. On the other
hand, we assume no experimental information on the heavier
$\Stop$-particle,~$\Stopz$, and on the mass of the $\cp$-odd
Higgs-boson, $\MA$. For large masses, $\Stopz$ and the heavy Higgs
bosons A, H and $\PH^\pm$ are very difficult to observe as a
consequence of background problems at the LHC and lacking energy at the LC.
For $\tan\be$ we assume that from measurements in the gaugino sector
either a fairly tight bound can be set if $\tan\be$ turns out to be
relatively small, e.g.\ $2.5 \leq \tan\be \leq 3.5$, or otherwise only a 
lower bound, e.g.\ $\tan\be \geq 10$ (see e.g.\ \citere{tbmeasurement}). 
As for the other parameters, the following values
are assumed, with uncertainties as expected from LHC~\cite{lhctdr} and
TESLA~\cite{teslatdr}:
$\mste = 500 \pm 2 \gev$,
$\sintt = -0.69 \pm 2\%$,
$A_{\Pb} = A_{\Pt} \pm 10\%$,
$\mu = -200 \pm 1 \gev$,
$M_2 = 400 \pm 2 \gev$  and
$\mgl = 500 \pm 10 \gev$.
In our analysis we assume a future uncertainty in the theoretical
prediction of $\mh$ of $\pm 0.5 \gev$.

\begin{figure}[htb]
\centerline{
\psfig{figure=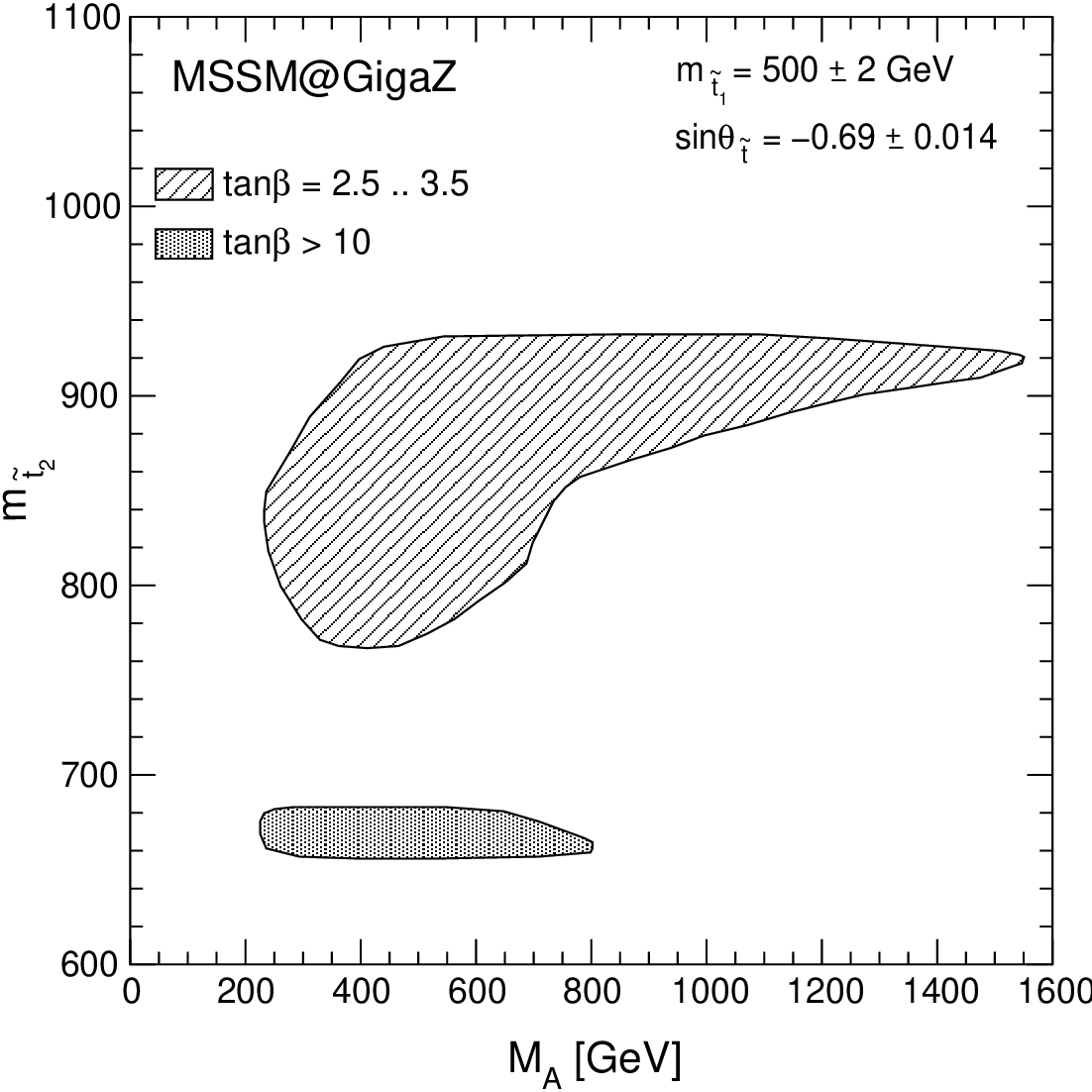,width=8cm,height=6.4cm}
}
\vspace{-6ex}
\caption{The region in the $\MA$--$\mstz$ plane, allowed by $1\,\si$ errors
obtained from the GigaZ measurements of $\MW$ and $\sweff$, taking as
hypothetical values
$\MW = 80.400 \pm 0.006$~GeV, $\sweff = 0.23140 \pm 0.00001$ and 
$\mh = 115.00 \pm 0.05$~GeV.
$\tb$ is assumed to be
experimentally constrained by $2.5 \leq \tb \leq 3.5$ or $\tb \geq 10$.
\label{fig:MSt2MA}
}
\vspace{-1em}
\end{figure}

In this scenario, assuming as experimental values
$\MW = 80.400 \pm 0.006$~GeV, $\sweff = 0.23140 \pm 0.00001$ and
$\mh = 115.00 \pm 0.05$~GeV, we find the allowed region in the 
$\MA$--$\mstz$ plane shown in \reffi{fig:MSt2MA}~\cite{gigaz}. 
In the case of large
$\tan\be$, where the prediction for $\mh$ depends only mildly on
$\tan\be$, the constraint from the $\mh$ measurement yields a rather
tight bound on $\mstz$~\cite{tampprec}. In combination with the information
from the data on $\MW$ and $\sweff$ this gives rise to a relatively small
allowed region in the $\MA$--$\mstz$ plane ($660 \gev \lsim \mstz \lsim
680 \gev$, $200 \gev \lsim \MA \lsim 800 \gev$). In the small $\tan\be$
region, on the other hand, where $\mh$ depends very sensitively on
$\tan\be$, a larger allowed region in the $\MA$--$\mstz$ plane is
obtained. In both cases, however, an upper bound on $\MA$ can be
established, which mainly arises from the precise measurement of 
$\sweff$. Note that we have assumed a measured value of $\sweff$
in accordance with the MSSM prediction in a region where the SUSY
particles do not completely decouple, i.e.\ a value slightly below the
corresponding SM prediction (see \reffi{fig:swmwsmmssm}). For the
experimental accuracy at an LC without the GigaZ mode no bound on $\MA$
could be inferred.

We thank J.~Erler, W.~Hollik and P.M.~Zerwas for collaboration on
various parts of the results presented here. G.W.\ thanks the organizers
of Loops and Legs 2000 for the invitation, the excellent organization
and the pleasant atmosphere during the workshop.

\end{document}